\documentclass[12pt]{iopart}

\usepackage{iopams}
\usepackage{siunitx}
\usepackage{booktabs}
\usepackage{multirow}
\usepackage{graphicx}
\usepackage{hyperref}
\usepackage[square,sort&compress,numbers]{natbib}
\renewcommand{\url}[1]{}

\newcommand{\ket}[1]{| #1 \rangle}
\newcommand{\abs}[1]{\left| #1 \right|}
\newcommand{\braket}[2]{\langle #1 | #2 \rangle}
\newcommand{\substack}[1]{\scalebox{0.7}{
    \renewcommand{\arraystretch}{0.5}
    \hspace{-3ex}
    $\begin{array}{c} #1 \end{array}$}} 

\begin{document}

\title{Self-guided tomography of time-frequency qudits}

\author{Laura Serino$^{1,*}$, Markus Rambach$^{2,3}$, Benjamin Brecht$^1$, Jacquiline Romero$^{2,3}$ and Christine Silberhorn$^1$}

\address{$^1$ Paderborn University, Integrated Quantum Optics, Institute for Photonic Quantum Systems (PhoQS), Warburgerstr.\ 100, 33098 Paderborn, Germany\\
$^2$ Australian Research Council Centre of Excellence for Engineered Quantum Systems, Brisbane, Queensland 4072, Australia\\ $^3$ School of Mathematics and Physics, University of Queensland, Brisbane, Queensland 4072, Australia}

\ead{$^*$ laura.serino@upb.de}

\begin{abstract}
High-dimensional time-frequency encodings have the potential to significantly advance quantum information science; however, practical applications require precise knowledge of the encoded quantum states, which becomes increasingly challenging for larger Hilbert spaces. Self-guided tomography (SGT) has emerged as a practical and scalable technique for this purpose in the spatial domain. Here, we apply SGT to estimate time-frequency states using a multi-output quantum pulse gate. We achieve fidelities of more than 99\% for 3- and 5-dimensional states without the need for calibration or post-processing. We demonstrate the robustness of SGT against statistical and environmental noise, highlighting its efficacy in the photon-starved regime typical of quantum information applications.
\end{abstract}

\section{Introduction}
The time-frequency degree of freedom of photons is gaining increasing attention in quantum information science due to its unique advantages \cite{brecht15, ansari18c, wang20, raymer20, lu23b}. It naturally supports high-dimensional encoding alphabets, allowing for the transmission of more information per photon compared to conventional binary systems \cite{leach12, wang20b}. This increased information density enhances resilience to noise and eavesdropping in quantum cryptography protocols and, therefore, makes high-dimensional encodings particularly advantageous for secure communication \cite{bechmann00, cerf02, sheridan10, islam17, ecker19, cozzolino19}. Additionally, the time-frequency domain offers greater resilience in transmission than the polarization and spatial degrees of freedom and is uniquely compatible with single-spatial-mode optical fibers, which are integral to modern telecommunication networks.

For practical applications of time-frequency encodings in quantum information science, precise knowledge of the encoded states is essential. In high dimensions, the process of characterizing quantum states---known as quantum state tomography---becomes inherently challenging due to the large associated Hilbert space. Namely, for a single qudit, the parameter space to be characterized grows as $d^2-1$, where $d$ is the dimension of the qudit \cite{banaszek99, thew02}. Standard quantum state tomography techniques, which apply the maximum-likelihood method to the results of an over-complete set of measurements \cite{hradil97, ansari17, bhattacharjee24}, and alternative interferometric methods \cite{wasilewski07, davis18, thiel20} become increasingly resource-expensive for large Hilbert spaces due to the significantly greater number of required measurements. Techniques such as compressive tomography \cite{gross10, ahn19, gil-lopez21a} require fewer measurements for an accurate quantum state estimation; however, the estimation algorithms are sensitive to noise and become more computationally demanding for high-dimensional systems. Moreover, in practical applications with imperfect detectors, many tomography techniques require an error calibration of the experimental setup to incorporate into the reconstruction algorithm and achieve reliable results \cite{cooper14, ansari17, serino23}.

Self-guided tomography (SGT) \cite{ferrie14} is an alternative technique that facilitates the estimation of a quantum state by maximizing its overlap with an iteratively updated guess without requiring any calibration or post-processing analysis. This method has been applied to the polarization \cite{chapman16} and spatial \cite{rambach21} degrees of freedom of photons, and has also been extended to perform quantum process tomography \cite{hou20}. In these applications, SGT has demonstrated scalability to high dimensions and resilience to statistical and environmental noise, which makes it particularly effective in the photon-starved regime.

In this work, we demonstrate the advantage of SGT applied to the time-frequency domain by accurately estimating a high-dimensional input state encoded in this degree of freedom. Using a so-called multi-output quantum pulse gate (mQPG) \cite{serino23}, we perform time-frequency projections and iteratively update the estimated state based on the results of their outcome. We achieve a fidelity to the encoded input state above $99\%$ in 3 and 5 dimensions, approaching the fidelity limit of the state preparation setup. Through SGT, we demonstrate the direct estimation of the quantum state without any calibration or post-processing, even in the presence of strong statistical and environmental noise.

\section{Method}

\begin{figure}
    \centering
    \includegraphics[width=\linewidth]{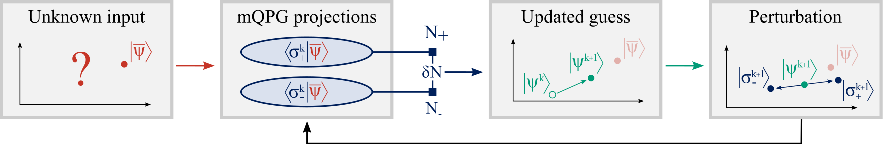}
    \caption{Schematic representation of the SGT procedure. The estimate $\ket{\psi^k}$ (initially a random guess) is iteratively updated based on the result of the projections of the true input state $\ket{\bar\psi}$ onto the two states $\ket{\sigma_\pm^k}$, obtained perturbing the latest estimate $\ket{\psi^k}$ in two opposite directions.}
    \label{fig:concept}
\end{figure}

The problem of quantum state tomography requires characterizing an unknown input state $\ket{\bar\psi}$ from a chosen Hilbert space. We choose a space described by temporal modes \cite{brecht15}, i.e., field-orthogonal wave-packet modes that encode information in the complex spectral amplitude of the electric field. Namely, we use the first $d$ Hermite-Gaussian modes $\{\ket{\alpha_j}\}_{j=1,..,d}$ as the computational basis of our $d$-dimensional Hilbert space. In this space, the input state can be expressed as $\ket{\bar\psi}=\sum_j{c_j\ket{\alpha_j}}$, where $c_j$ are complex coefficients. We note that this description intrinsically assumes a single-photon state in input and focuses solely on the temporal-mode structure of the state. We assume pure input states, as in the original formulation of SGT \cite{ferrie14}, although this technique can also be adapted to mixed states if necessary \cite{rambach21}.

To implement SGT, we begin by generating a set of random complex coefficients $c_j^0$ to provide an initial guess of the target state: $\ket{\psi^0}=\sum_j{c_j^0\ket{\alpha_j}}$.
At each iteration $k$ of the algorithm, this estimate will be updated to a more accurate $\ket{\psi^k}=\sum_j{c_j^k\ket{\alpha_j}}$ based on the result of the time-frequency projections performed by the mQPG (\fref{fig:concept}). 

The mQPG \cite{serino23} serves as a mode-sorter for single-photon-level time-frequency states. Its working principle is based on sum-frequency generation in a dispersion-engineered nonlinear waveguide driven by a spectrally shaped pump pulse \cite{brecht14}. The mQPG operation is described by a transfer function, which is the product of the energy conservation condition, determined by the pump function, and the phase-matching function, describing momentum conservation (see inset in \fref{fig:setup}). By choosing a pump wavelength that is group-velocity-matched to the input wavelength in the nonlinear medium, we achieve a horizontal phase-matching function that facilitates mode-selective operation in each channel of the mQPG \cite{eckstein11, ansari17}. Thus, the probability of upconversion in each channel is proportional to the complex spectral overlap between the input mode $\ket{\bar\psi}$ and the pump mode $\ket\sigma$, which can be selected via spectral shaping. Effectively, the mQPG projects a high-dimensional input state onto the selected pump modes and yields the result of each projection into a separate output channel, corresponding to a distinct output frequency that can be read out using a spectrograph. 

The states for the projections at each iteration $k$ are chosen starting from the most updated guess $\ket{\psi^k}$. We perturb each complex coefficient $c_j^k$ in a random direction $(\Delta_k)_j\in \{1, -1, i, -i\}$ \cite{utreras19} with strength $\beta_k=b/(k+1)^t$, where $b$ and $t$ are hyperparameters of the algorithm, optimised once by trial and error in simulations. From this perturbation, we obtain the two states $\ket{\sigma_\pm^k} = \ket{\psi^k \pm \beta_k\Delta_k}$, and we assign each to a channel of a two-output mQPG. 
The mQPG projects $\ket{\bar\psi}$ onto $\ket{\sigma_+^k}$ and $\ket{\sigma_-^k}$ and yields the results of the projections as $N_+$ and $N_-$ clicks detected in the respective channels. 

From the pseudo-normalized quantity $\delta\!N = (N_+ - N_-)/(N_+ + N_-)$, we calculate the gradient $g_k=\delta\!N \Delta_k / 2 \beta_k$ which indicates the magnitude and direction of the distance vector between the target state $\ket{\bar\psi}$ and its estimate $\ket{\psi^k}$.
We use this gradient to compute the next estimate in the iteration as $\ket{\psi^{k+1}} = \ket{\psi^k+\alpha_k g_k}$, where $\alpha_k=a/(k+1+A)^s$ determines the step size in the direction of the gradient, with $a$, $A$ and $s$ algorithm hyperparameters. 

This iterative procedure is repeated for a chosen number of steps $K$, after which we obtain the final estimation $\ket{\psi^K}$. 
We then calculate the infidelity $1-\abs{\braket{\psi^K}{\bar\psi}}^2$ to quantify the residual distance between the reconstructed state $\ket{\psi^K}$ and the target state $\ket{\bar{\psi}}$.

\section{Experiment}

\begin{figure}
    \centering
    \includegraphics{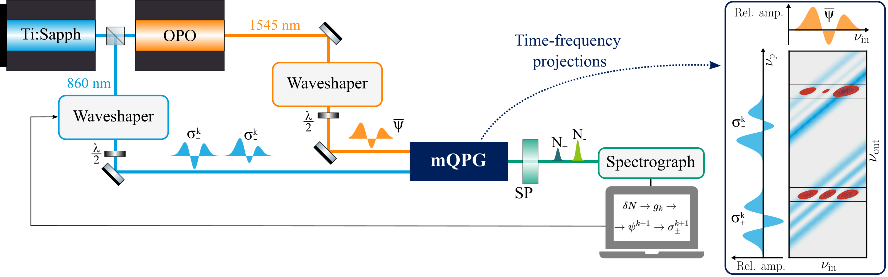}
    \caption{Schematic of the experimental setup. An optical parametric oscillator (OPO) system driven by an ultrafast Ti:Sapph laser produces the input (orange) and pump (blue) pulses. The input state $\ket{\bar\psi}$ is generated by a commercial waveshaper (Finisar 4000S), whereas the pump pulse is shaped by an in-house-built folded 4-f waveshaper \cite{monmayrant10} to generate $\ket{\sigma_\pm^k}$. Using two half-waveplates, we align the polarization of input and pump to horizontal and vertical, respectively, as required by the mQPG process. The mQPG projects $\ket{\bar\psi}$ onto $\ket{\sigma_\pm^k}$, yielding the results as $N_+$ and $N_-$ upconverted photons in the two output channels. The inset shows the frequency-space representation of the mQPG projections, described by a transfer function (red), which is the product of the phase-matching function (contoured by horizontal black lines) and the energy conservation condition determined by the pump spectrum (blue). A CCD spectrograph (Andor Shamrock 500i) detects the output photons, discriminating the two output frequencies. From the pseudo-normalized difference in counts $\delta N$, we calculate the gradient $g_k$ to determine the next estimation $\ket{\psi^{k+1}}$ which, in turn, is the starting point for $\ket{\sigma_\pm^{k+1}}$ in the following iteration.}
    \label{fig:setup}
\end{figure}

\Fref{fig:setup} shows a schematic of the experimental setup.
Input and pump pulses are generated by an optical parametric oscillator (OPO) system driven by an ultrafast Ti:Sapph laser emitting \SI{150}{\femto\second} coherent pulses at a repetition rate of \SI{80}{\mega\hertz}. The input pulse, centred at \SI{1545}{\nano\meter}, is shaped by a commercial waveshaper (Finisar 4000S) into a random superposition $\ket{\bar\psi}$ of the first $d$ Hermite-Gaussian functions, with $d\in\{3,5\}$ dimensionality of the Hilbert space.

The first estimate $\ket{\psi^0}$ of the input state is chosen randomly, and its coefficients are perturbed in two opposite directions to find the two states $\ket{\sigma_\pm^0}$ used for the mQPG projections. The pump pulse, centred at \SI{860}{\nano\meter}, is shaped by an in-house-built folded 4-f waveshaper \cite{weiner00, monmayrant10} to generate the complex spectra of $\ket{\sigma_+^0}$ and $\ket{\sigma_-^0}$. This custom-built waveshaper is necessary because commercial waveshapers are not yet available for this wavelength range.

The mQPG waveguide used in this experiment is realized in titanium in-diffused lithium niobate operated at \SI{160}{\celsius}, with a periodic poling pattern consisting of an alternation of unpoled regions and regions which are poled with a period of \SI{4.32}{\micro\meter} \cite{serino23}. This pattern enables the mQPG to perform time-frequency projections in two different channels centred at two distinct frequencies. The spectra of $\ket{\sigma_\pm^0}$ in the pump pulse, therefore, are centred at two offset frequencies matched to the two output frequencies of the mQPG.

Input and pump pulses are coupled into the mQPG waveguide, which projects the input state $\ket{\bar\psi}$ onto $\ket{\sigma_+^0}$ and $\ket{\sigma_-^0}$. Each copy of the input state (i.e., each input photon) is upconverted in the ``+'' or ``-'' channel with probability proportional to the overlap $\abs{\braket{\sigma_\pm^0}{\bar\psi}}^2$. 
The upconverted photons are detected by a single-photon-sensitive spectrograph (Andor Shamrock 500i), which discriminates the output frequencies and integrates over multiple pulses, providing the total counts $N_+$ and $N_-$ observed in each channel during the integration period.

From these counts, we calculate the first gradient $g_0$ to determine the next estimate $\ket{\psi^1}$. This procedure is repeated for each iteration $k$, refining the estimate $\ket{\psi^k}$ closer to the true input state $\ket{\bar\psi}$. We perform a total of $K=200$ iterations in $d=3$ and $K=300$ in $d=5$.

For each dimensionality ($d=3$ and $d=5$), we repeated the SGT procedure on 100 different input states chosen randomly with Haar measure \cite{mezzadri07}.
We tested this method on the same set of input states in different conditions of statistical noise quantified by $\sqrt{N}/N$, where $N$ denotes the number of maximum counts in each channel when the input and pump states perfectly overlap.
The values $N=10^2$, $N=10^3$ and $N=10^4$ were achieved by adjusting the integration time of the spectrograph from \SI{1}{\ms} to \SI{10}{\ms} and \SI{100}{\ms}, corresponding to approximately \SI{2.5e-3}{} clicks per pulse; $N=10^5$ was obtained with an integration time of \SI{100}{\ms} and a larger photon number in the input, resulting in approximately \SI{2.5e-2}{} clicks per pulse. Additionally, we performed standard tomography based on maximum-likelihood estimation \cite{ansari17} with a single channel of the mQPG under identical noise conditions and with the same input states for a comparative analysis.

We note that the spectrograph output was strongly affected by electronic read-out noise with a mean of 890 counts and a standard deviation $\sigma=14$ in each channel. This was the dominant source of environmental noise during the measurements and was independent of the integration time. To limit its impact, we subtracted the minimum ``constant'' background value of 820, chosen at a distance of $5\sigma$ from the mean to prevent negative count artifacts. This correction left a residual background of 70 counts per channel with the same standard deviation $\sigma$ as the main source of environmental noise.

To complement the experimental data, we performed realistic simulations of the SGT process taking into account the measured electronic noise and the imperfect mQPG projections. We used these simulations to find the set of hyperparameters $(b, t, a, A, s)$ that optimized the convergence of the estimated quantum state. We then fine-tuned these values in the experiment, observing how they affected the convergence rate of the infidelity for the same input state.

\section{Results and discussion}

\begin{figure}
    \centering
    \includegraphics[width=15cm]{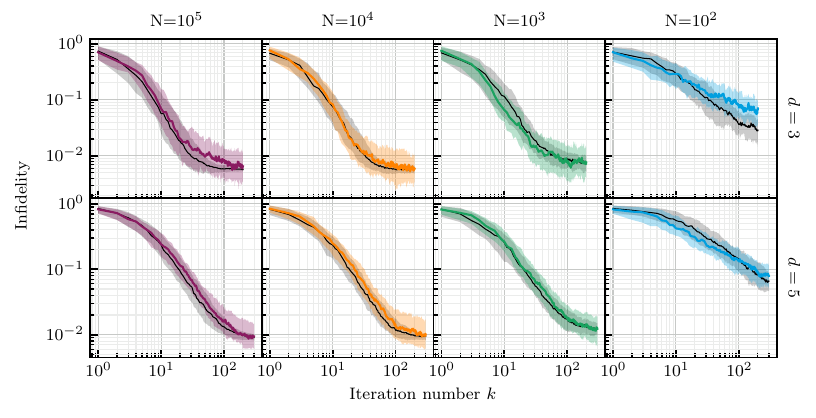}
    \caption{The coloured lines show the experimental infidelities of the reconstructed states at each iteration of the SGT algorithm for different values $N$ of the maximum counts per iteration in each channel. The lines represent the median infidelity of the complete population of 100 different input states, whereas the shaded area shows the upper and lower quartile of the infidelities $(50\pm25)\%$. In close agreement with the experiment, the black lines show the results of simulations that take into account experimental imperfections.}
    \label{fig:results_sep}
\end{figure}

\begin{figure}
    \centering
    \includegraphics{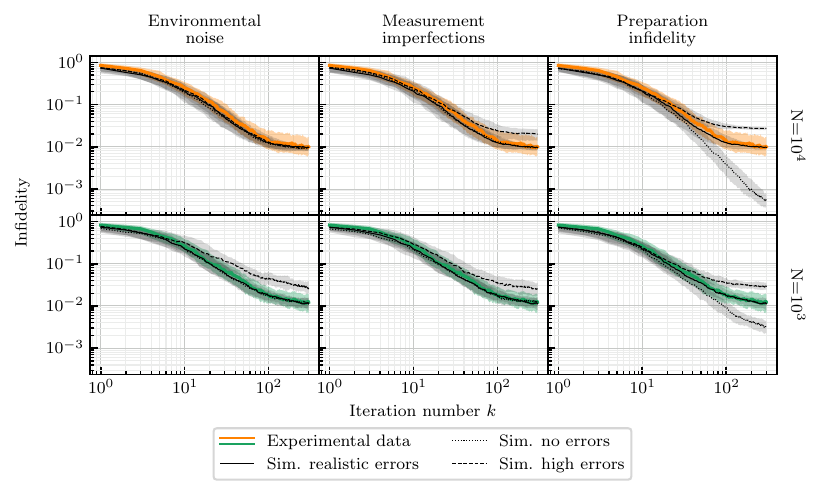}
    \caption{Effect of different sources of imperfections on the SGT convergence curve based on the same set of 100 input states in $d=5$ for the two statistics $N=10^3$ and $N=10^4$. The experimental data (coloured line) and realistic simulation (solid black line) are the same as shown in \fref{fig:results_sep} and show very good overlap, indicating a good understanding of the experimental errors in the system. In each column, we change the value of a particular type of error in the simulations: the dotted lines show what happens when we completely remove that source of error, whereas the dashed lines show the effect of an increase in that error source (10 times greater environmental noise and measurement imperfections, 3 times larger preparation infidelity). The dotted line is clearly visible only in the rightmost plots, suggesting that preparation infidelity is likely the major limiting factor in the quality of the reconstructed states. See the text for more information.}
    \label{fig:noise_sim2}
\end{figure}

\begin{table}
    \centering
    \begin{indented}
    \item[]\begin{tabular}{c|cc|cc}
        \toprule
        & \multicolumn{2}{c|}{$\mathbf{d=3}$} & \multicolumn{2}{c}{$\mathbf{d=5}$}\\
        $N$ & SGT & MLST & SGT & MLST \\
        \midrule
        $10^5$ & $99.35\substack{+0.29 \\ -0.51}\%$ & $99.12\substack{+0.47 \\ -0.58}\%$ & $99.06\substack{+0.33 \\ -0.54}\%$ & $96.76\substack{+0.52 \\ -1.02}\%$\\
        $10^4$ & $99.40\substack{+0.24 \\ -0.38}\%$ & $99.22\substack{+0.49 \\ -0.34}\%$ & $98.99\substack{+0.36 \\ -0.71}\%$ & $97.33\substack{+0.72 \\ -0.79}\%$\\
        $10^3$ & $99.24\substack{+0.35 \\ -0.55}\%$ & $99.52\substack{+0.22 \\ -0.88}\%$ & $98.77\substack{+0.55 \\ -0.62}\%$ & $96.14\substack{+1.02 \\ -0.90}\%$\\
        $10^2$ & $93.0\substack{+3.4 \\ -3.0}\%$ & $86.9\substack{+6.0 \\ -8.0}\%$ & $92.1\substack{+3.0 \\ -3.7}\%$ & $60.4\substack{+5.3 \\ -4.7}\%$\\
        \bottomrule
    \end{tabular}
    \end{indented}
    \caption{Comparison of the final fidelity achieved through SGT and through maximum-likelihood state tomography (MLST) \cite{ansari17} performed with the same experimental setup in identical environmental conditions. The values indicate the median value over a set of 100 random input states, and the error shows the upper and lower quartile values of the distribution. The same set of input states was characterized with both methods.}
    \label{tab:fidelity}
\end{table}

\Fref{fig:results_sep} shows (in colour) the median infidelity of the reconstructed states at each iteration of the process. 
For high photon numbers, we reach 90\% fidelity after only 10 iterations in $d=3$ and 20 iterations in $d=5$. These values are similar to what Rambach \textit{et al.}\ \cite{rambach21} obtained in the spatial domain for the same dimensionalities, showcasing the adaptability of SGT to different experimental implementations. 

However, after this value, the decrease in infidelity slows down until it reaches a plateau. This is in contrast to the results achieved in the spatial domain, where the infidelity continued to decrease indefinitely within the measured number of iterations. 
Notably, tuning the hyperparameters only affects the convergence speed but not the final infidelity.

The saturation of the infidelity can be attributed to a combination of two factors: systematic errors of the measurement device and imperfections in the preparation of the input states. 
Although we assumed ideal input preparation when calculating the fidelity of the estimated states, an infidelity of 1\% in the waveshaping system that generates the input states can significantly decrease the maximum precision achievable by SGT.
Additionally, although this technique is highly resilient to statistical noise due to its iterative and randomised component, it can still be affected by systematic noise from imperfect mQPG operation \cite{ansari17, gil-lopez21, serino23}.

The black lines in \fref{fig:results_sep} show the simulated results of the SGT process based on the measured electronic noise and mQPG imperfections (quantified as average cross-talk between orthogonal states \cite{serino23}), in addition to an estimated infidelity in the input preparation of $(0.6\pm0.1)\%$ in $d=3$ and $(0.9\pm0.1)\%$ in $d=5$. We find excellent agreement between the simulation results and the experimental infidelity curve, visible both in the convergence rate and in the saturation value of the infidelity, thus supporting the assumptions on the imperfections in the input preparation.

\Fref{fig:noise_sim2} illustrates the isolated effects of different types of imperfections (environmental noise, systematic errors in the measurement device and preparation infidelity) on the two datasets in $d=5$ with $N=10^3$ and $N=10^4$. We always show the experimental data and realistic simulation for comparison as coloured and black solid lines, respectively. In each column, we study the effect of each particular type of imperfection by changing its value in the simulations. The dotted lines are the results of simulations in which we completely remove that source of error; the only visible improvement is observed when the preparation infidelity is eliminated (last column in Fig.~\ref{fig:noise_sim2}). The dashed lines show the effect of an increase in the selected error source (10 times greater environmental noise, 10 times more systematic measurement imperfections, and 3 times larger preparation infidelity), keeping the others unchanged. While this leads to a larger final infidelity in general, the dataset with fewer counts is significantly more affected by an increase in the environmental noise. Overall, one can notice that the limited preparation quality of the input states represents the most significant constraint to the maximum achievable fidelity in the current experimental conditions.

Despite this technical limitation, SGT consistently achieves higher fidelity than standard maximum-likelihood tomography \cite{ansari17} performed with the same experimental setup in identical environmental conditions (see \tref{tab:fidelity} for detailed comparisons). Notably, the estimate of the input state is known in real-time at every step of the process without requiring post-processing.

Furthermore, SGT demonstrates superior performance even under low count rates. In both dimensionalities, we achieve fidelities of approximately 99\% with only $10^3$ counts per measurement. Even with as few as 100 counts per measurement, the fidelity improves with each iteration, albeit more slowly, reaching 90\% after approximately 100 iterations in $d=3$ and 200 iterations in $d=5$. This resilience to statistical noise highlights SGT as an optimal method for quantum state tomography in the photon-starved regime.

\section{Conclusion}
We applied SGT for the first time to time-frequency qudits, showcasing the versatility of this method in different degrees of freedom. We achieved a fidelity of the estimated states above 99\% in 3 and 5 dimensions without the need for calibration or post-processing. The experimental results highlight the resilience of this technique to statistical and environmental noise, favoured by its iterative character. Through realistic simulations that closely reproduce the experimental data, we infer that the ultimate fidelity of our estimates is predominantly limited by the accuracy of the input state preparation. These results demonstrate the robust and adaptable nature of this technique, paving the way for further exploration and potential applications in quantum information science.

\ack
The authors would like to thank C.\ Ferrie for helpful and inspiring discussions. 
LS has received funding from the European Union’s Horizon Europe research and innovation programme under grant agreement No 899587 (STORMYTUNE). JR and MR are supported by Australian Research Council Centre of Excellence for Engineered Quantum Systems (EQUS, CE170100009).

\small
\bibliographystyle{iopart-num}
\bibliography{bibliography}

\end{document}